\documentclass[sigconf]{acmart}
\makeatletter
\def\@ACM@checkaffil{%
    \if@ACM@instpresent\else
    \ClassWarningNoLine{\@classname}{No institution present for an affiliation}%
    \fi
    \if@ACM@citypresent\else
    \ClassWarningNoLine{\@classname}{No city present for an affiliation}%
    \fi
    \if@ACM@countrypresent\else
        \ClassWarningNoLine{\@classname}{No country present for an affiliation}%
    \fi
}
\makeatother
\AtBeginDocument{%
  \providecommand\BibTeX{{%
    \normalfont B\kern-0.5em{\scshape i\kern-0.25em b}\kern-0.8em\TeX}}}

\copyrightyear{2024}
\acmYear{2024}
\setcopyright{rightsretained}
\acmConference[CAIN 2024]{Conference on AI Engineering Software Engineering
for AI}{April 14--15, 2024}{Lisbon, Portugal}
\acmBooktitle{Conference on AI Engineering Software Engineering for AI (CAIN
2024), April 14--15, 2024, Lisbon,
Portugal}
\acmDOI{10.1145/3644815.3644950}
\acmISBN{979-8-4007-0591-5/24/04}

\usepackage{makecell}
\usepackage[htt]{hyphenat}
\usepackage{listings}
\usepackage{multirow}
\usepackage{booktabs}

\begin{document}

\title{(Why) Is My Prompt Getting Worse? \\ Rethinking Regression Testing for Evolving LLM APIs}

\author{Wanqin Ma}
\email{wmaag@connect.ust.hk}
\authornote{These two authors contribute equally to the work.}
\affiliation{%
  \institution{The Hong Kong University of Science and Technology}
}

\author{Chenyang Yang}
\email{cyang3@cs.cmu.edu}
\authornotemark[1]
\affiliation{%
  \institution{Carnegie Mellon University}
}

\author{Christian K{\"a}stner}
\affiliation{%
  \institution{Carnegie Mellon University}
}

\renewcommand{\shortauthors}{Ma et al.}

\newcommand{\nbc}[3]{
 {\colorbox{#3}{\bfseries\sffamily\scriptsize\textcolor{white}{#1}}}
 {\textcolor{#3}{\sf\small$\blacktriangleright$\textit{#2}$\blacktriangleleft$}}
 }
\newcommand{\cyang}[1]{\nbc{CY}{#1}{teal}}

\lstset{
  language=Python,
  keywords={},
  commentstyle=\textbf,
  stringstyle=\text,
  extendedchars=false,
  basicstyle=\ttfamily,
  columns=fullflexible,
  frame=single,
  breaklines=true,
  breakatwhitespace=true,
  breakindent=2\dimen0,
}
\newcommand{\hl}[1]{#1}

\begin{abstract}
Large Language Models (LLMs) are increasingly integrated into software applications.
Downstream application developers often access LLMs through APIs provided as a service.
However, LLM APIs are often updated silently and scheduled to be deprecated, forcing users to continuously adapt to evolving models.
This can cause performance regression and affect prompt design choices, as evidenced by our case study on toxicity detection.
Based on our case study, we emphasize the need for and re-examine the concept of regression testing for evolving LLM APIs.
We argue that regression testing LLMs requires fundamental changes to traditional testing approaches, due to different correctness notions, prompting brittleness, and non-determinism in LLM APIs.

\end{abstract}



\maketitle

\section{Introduction}

\begin{figure}
    \centering
    \includegraphics[width=0.95\linewidth]{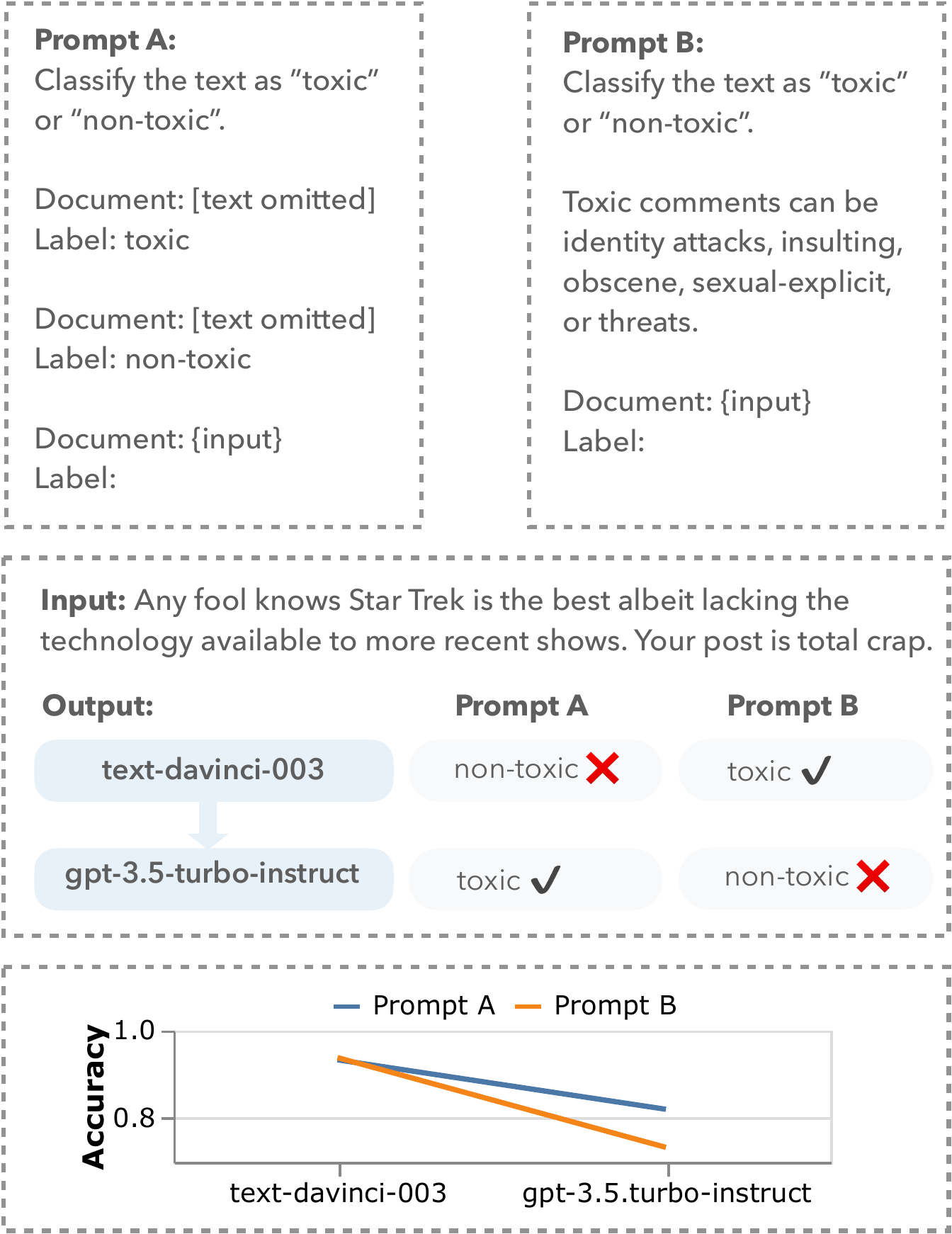}
    \caption{An LLM API update from \texttt{text-davinci-003} to \texttt{gpt-3.5-turbo-instruct} causes a major performance downgrade on classifying toxic comments. The API update also changes the prompt choice: Prompt A (left) now outperforms Prompt B (right) by 8.7\% accuracy.}
    \label{fig:mot_ex}
    \vspace{-10pt}
\end{figure}

Large Language Models (LLMs) are increasingly integrated into software applications~\cite{kaddour2023challenges}.
Due to the high cost of developing and maintaining in-house LLMs, many applications rely on LLM APIs provided by companies like OpenAI, Anthropic, and Google~\cite{ChatGPT, Bard, Claude}.
Although LLM APIs provide easy access to state-of-the-art models, they also bring in uncertainties for their downstream applications:
It is not uncommon for application developers to find their carefully engineered prompts that worked yesterday work less well after updates from the LLM provider's side~\cite{prompt-not-work, chen2023chatgpts}.
In Figure~\ref{fig:mot_ex}, we highlight such an example from our case study on a toxicity detection task, where the LLM API update from \texttt{text-davinci-003} to \texttt{gpt-3.5-turbo-instruct} causes a major performance downgrade and changes the good choices for prompt selection.

Similar to traditional web service API~\cite{li2013does} and more conventional ML APIs~\cite{EvolvingAIAPI}, updates to server-side LLM API controlled by a different party are hard to deal with.
First, LLM APIs can be updated silently: OpenAI's \texttt{gpt-3.5-turbo} model has been updated twice \hl{(by Nov 2023)} but the updates are not visible to the downstream developers. 
\hl{Such silent API updates change only the underlying LLM but not the API signature}\footnote{\hl{For our purpose, an LLM API describes both the signature of the service and its behavior. In this work, we primarily focus on behavior changes that are difficult to document and detect.}}, causing unexpected behavioral changes (e.g., formatting of generated code)~\cite{chen2023chatgpts} \hl{to the application developers}.
Second, LLM APIs are scheduled to be deprecated and discontinued~\cite{deprecation}, effectively forcing application developers to adopt newer API versions.
For example, the \texttt{text-davinci-003} model will be deprecated on Jan 2024. The forced transition to \texttt{gpt-3.5-turbo-instruct} can cause unexpected prompt performance changes, including the introduced performance downgrade as illustrated in our example in Figure~\ref{fig:mot_ex}.

To cope with evolving LLM APIs, application developers need support for monitoring and analyzing how their prompts perform differently when the LLM API changes.
Existing software engineering practices suggest that regression testing is essential for identifying changes between software versions, often particularly to ensure that fixed bugs are not reintroduced~\cite{SEbook}.
We argue that LLM application developers should take a similar approach.
However, existing regression testing practices can not directly translate in the LLM context, as we will illustrate.
Based on our observations in the case study, we highlight three fundamental changes for regression testing LLM APIs:

First, LLM regression tests should be defined at a different \textit{granularity}.
In traditional software engineering, a single breaking regression test would indicate a bug in the software implementation.
In contrast, it is common for ML models to change predictions for individual data points after updates.
The common practice is to examine overall model accuracy, which has been criticized for being coarse-grained~\cite{ribeiro-etal-2020-beyond}.
To gain a more nuanced understanding than overall model accuracy, LLM regression tests should be defined over data \textit{slices} rather than on single predictions or the entire dataset.
This calls for a different \textit{correctness} notion, as ``regression'' is defined over slice-level aggregated metrics and the slice-level test only fails when the metrics change beyond a threshold.

Second, LLM regression tests need to monitor both model and \textit{prompt} updates.
It is well-known that prompt engineering can greatly influence LLMs' performance~\cite{prompt_survey}.
As we will show, we observed that different prompt designs regress or improve differently on the same API update,
making the optimal prompt design change from API version to version.
We argue that tracking both LLM and prompt \textit{versions} is essential for LLM regression tests.

Third, LLM regression tests need to deal with \textit{non-determinism} of LLM APIs.
LLMs are known to produce non-deterministic outputs: Non-determinism is often introduced intentionally for generating high-quality outputs with a non-zero temperature~\cite[e.g.,][]{fried2023incoder}, but can even be observed with a zero temperature setting~\cite{ouyang2023llm}, \hl{where the LLM should deterministically predict the next most likely token.}
It is necessary to deal with \textit{flakiness} in LLM regression tests by considering their inherent non-determinism.

In summary, our paper has the following contributions:
\begin{itemize}
    \item An exploratory case study on toxicity detection with the GPT-3.5 model family, showing API upgrades can cause significant performance deterioration, and that prompt is an important factor in behavioral changes.
    \item A re-examination of the concept of regression testing for LLM APIs and its required fundamental changes, due to different correctness notions, prompting brittleness, and non-determinism in LLMs.
    \item A vision on research opportunities in supporting systematic regression testing for prompting LLM APIs.
\end{itemize}

\begin{table*}[t]
\centering
\small
\begin{tabular}{llll}
\toprule
\textbf{Model} & \textbf{Endpoint Type} & \textbf{Training Method} & \textbf{Release Date} \\
\midrule
\texttt{gpt-3.5-turbo-instruct}              & Text completion & RLHF & Sep 2023  \\
\texttt{gpt-3.5-turbo-0613}                  & Chat & RLHF & June 2023  \\
\texttt{gpt-3.5-turbo-0301}                  & Chat & RLHF & Mar 2023 \\
\texttt{text-davinci-003}                    & Text completion & RLHF & Nov 2022 \\
\texttt{text-davinci-002}                    & Text completion & fine-tuning & Mar 2022 \\
\bottomrule
\end{tabular}
\caption{Representative models from OpenAI's GPT-3.5 family~\cite{GPT3.5}, sorted by release date.}
\label{tab:models}
\vspace{-20pt}
\end{table*}

\section{Background and Related Work}
\subsection{Evolving AI APIs}
As ML models are increasingly provided as a cloud service through APIs (e.g., Perspective API~\cite{Perspective}, ChatGPT~\cite{ChatGPT}, Amazon Rekognition~\cite{Mishra_2019}),
it has been noticed that these models evolve over time without clear documentation~\cite{8919078, tu2023chatlog}, similar to traditional web service API~\cite{li2013does}.
This can pose risks to downstream application developers, who do not have control over model updates and can potentially suffer from performance regression~\cite{chen2023chatgpts, EvolvingAIAPI}.

Beyond demonstrating the problem, there has only been limited work on actually supporting developers facing evolving APIs not under their control. 
The most prominent example for ML models is done by~\citet{EvolvingAIAPI}, where they focus on detecting changes in the label space and prediction confidence for vision APIs.
Our work extends the existing literature by explicitly adapting the concept of regression testing in the LLM contexts and highlighting the need for more nuanced regression test suites.

\subsection{The Rise of Prompting LLMs}
\looseness=-1
LLMs present a fundamental shift in NLP applications through the \textit{prompting} interface, which allows rapid prototyping and iterations~\cite{prompt_survey}:
Application developers can easily tweak prompts and validate prompts on a few examples without the need to curate data and build models.
In a sense, the LLM together with a specific prompt can be considered equivalent to a traditional specifically-trained ML model for a specific task, such as toxicity detection.
However, the \textit{prompting} paradigm also brings in the risk of prompt brittleness,
as prompts can be sensitive to small changes~\cite{lu-etal-2022-fantastically} and the good choices for prompts change when the LLM changes.
Our work highlights prompts as an additional factor to consider for regression testing LLMs.

\subsection{ML Model Testing}
ML models are usually evaluated by model fit using aggregated metrics like accuracy,
as models are expected to make occasional mistakes~\cite{SEAIbook}.
However, traditional model evaluation has been criticized for being coarse-grained~\cite{ribeiro-etal-2020-beyond} and suffering from issues like spurious correlations~\cite{spurious}.
Therefore, recent work has proposed nuanced behavioral model testing as an alternative~\cite{naik-etal-2018-stress, ribeiro-etal-2020-beyond}, where the testers explore nuanced model behaviors beyond a single score.

Prior work has explored different methods to explore and test model behaviors~\cite[e.g.,][]{ribeiro-etal-2020-beyond, ribeiro-lundberg-2022-adaptive, yang2023beyond}, as well as different ways to automate testing specific model behaviors~\cite[e.g.,][]{sun20translation, sun22isotopic} (see~\citet{yang23SafeAI} for a detailed survey).
Another line of work on data slicing~\cite[e.g.,][]{zeno, eyuboglu2022domino, fse19slicing} focuses on identifying data regions where a model under-performs.
Our work introduces a new scenario for ML model testing: \textit{regression testing} over evolving LLM APIs.

\section{Case Study: Toxicity Detection}
\label{sec:case-study}
Since regression of evolving LLM APIs is an emerging problem of which we have little understanding, 
we first explored the problem with an exploratory case study.
We picked a paradigmatic case~\cite{yin2009case} of toxicity detection, a task widely used for online content moderation and long performed by models specifically trained for that task~\cite{hosseini2017deceiving}, but recently LLMs with a suitable prompt have shown similar or better performance~\cite{wang2022toxicity}.
Our case study aims to explore (a) how prompt behaviors change (regress) over LLM updates and (b) where regressions can be detected.%
\footnote{\hl{Code available at} \url{https://github.com/MAWanqin2002/LLM_Regression_Testing}.}

\subsection{Experiment Setup}
\subsubsection{Datasets.}
We selected two toxicity detection datasets for our case study: Civil Comments~\cite{jigsaw-unintended-bias-in-toxicity-classification} and GitHub Discussion~\cite{Github},
covering different contents (\textit{generic} vs. \textit{specialized}) and text lengths (\textit{short} vs. \textit{long}) for toxicity detection.

The Civil Comments dataset is collected from Civil Comments platform, representing a wide range of comments on the Internet.
We sampled 1000 comments from the dataset, among which 41 are toxic and 959 are non-toxic.

The GitHub Discussion Dataset contains 174 discussions, among which 74 are toxic and 100 are non-toxic. 
The 74 toxic discussions are collected using the links provided by an existing study~\cite{Github}, and we randomly sample another 100 non-toxic discussions from GitHub.

\subsubsection{Models.}
We selected five widely used models from OpenAI's GPT-3.5 family~\cite{GPT3.5}: %
\texttt{gpt-3.5-turbo-instruct}, \texttt{gpt-3.5-turbo-0613}, \texttt{gpt-3.5-turbo-0301}, \texttt{text-davinci-003}, and \texttt{text-davi\-nci-002} (shown in Table~\ref{tab:models}).
These models were released over a span of only 18 months, from March 2022 to September 2023, covering different endpoint types (chat vs. completion) and training methods (fine-tuning vs. RLHF).
We treat each model pair as a potential update and study 10 model update pairs in our experiment.

Noticeably, four out of these five models are already scheduled to be deprecated in 2024, effectively forcing application developers to switch to one of the newer models.
The models are also updated silently: \texttt{gpt-3.5-turbo-0301} and \texttt{gpt-3.5-turbo-0613} are snapshots of the \texttt{gpt-3.5-turbo} model, which will soon point to \texttt{gpt-3.5-turbo-1106}.

\subsubsection{Prompts.}
We employed four prompting strategies to explore how they behave differently on model updates:
\begin{itemize}
    \item \textbf{Simple instruction (P1)}: The prompt instructs the model to classify the text as ``toxic'' or ``non-toxic'', followed by the text to classify. This serves as a simple baseline a developer might first try.
    \item \textbf{Simple instruction, placed last (P2)}: The same as above but put instructions after the text. This design follows the insight that LLMs have recency bias~\cite{pmlr-v139-zhao21c} and stating instructions last makes LLMs less likely to ramble~\cite{Liu2023InstructionPM}.
    \item \textbf{Detailed instruction (P3)}: The prompt first describes the classification goal in detail, explaining what the developer deems toxic. The description is followed by the classification instruction and the text to classify.
    \item \textbf{Simple instruction + Few-shot examples (P4)}: After simple instructions, the prompt shows two examples, one toxic and the other non-toxic, followed by the text to classify. This follows the popular in-context learning paradigm~\cite{NEURIPS2020_1457c0d6}.
\end{itemize}

We share the prompt templates in Figure~\ref{tab_prompt}.

\begin{figure}[t]
\small
\begin{lstlisting}
# Simple instruction (P1)
Classify the GitHub discussion as "toxic" or "non-toxic". Only reply with the label.
Document: {text}

# Simple instruction, placed last (P2)
Document: {text}
Classify the GitHub discussion as "toxic" or "non-toxic". Only reply with the label.

# Detailed instruction (P3)
Below is a GitHub discussion. Sometimes the discussion can get heated and have toxic comments. Toxic comments can contain curse words, can sound condescending, can be mean to others, or can make people feel angry without using offensive words.
Classify the GitHub discussion as "toxic" or "non-toxic". Only reply with the label.
Document: {text}

# Simple instruction + Few-shot examples (P4)
Classify the GitHub discussion as "toxic" or "non-toxic". Only reply with the label.

Document: [text omitted]
Label: toxic

Document: [text omitted]
Label: non-toxic

Document: {text}
Label:
\end{lstlisting}

\vspace{-10pt}
\caption{Prompt templates for our experiments on the GitHub discussion dataset. Templates for the Civil Comments dataset are similar with some adaptations.}
\label{tab_prompt}
\vspace{-10pt}
\end{figure}

\subsubsection{Metrics.}
To evaluate the accuracy of each model + prompt combination and monitor their changes, we use the standard performance metrics accuracy and F1, and set model temperature to 0 to obtain the most likely predictions.

\begin{table*}[ht]
  \small
  \begin{tabular}{lrrrrrrrr}
    \toprule
    \multirow{2}{*}{\textbf{Model}}&\multicolumn{4}{c}{\textbf{Civil Comments}} & \multicolumn{4}{c}{\textbf{GitHub Discussion}} \\    
    \cmidrule(lr){2-5}\cmidrule(lr){6-9}   
            &\textbf{P1}&\textbf{P2}&\textbf{P3}&\textbf{P4} &\textbf{P1}&\textbf{P2}&\textbf{P3}&\textbf{P4}

\\
    \midrule
    \texttt{gpt-3.5-turbo-instruct}& 0.688 & {0.518} &0.733 & {\textbf{0.820}} & 0.638 & \textbf{0.793} & 0.770 & \textbf{0.793}\\
    \texttt{gpt-3.5-turbo-0613} & {0.671}& 0.745 & {\textbf{0.928}} & 0.774 & 0.822 & \textbf{0.862} & \textbf{0.856} & 0.776\\
    \texttt{gpt-3.5-turbo-0301}&0.767 & 0.743 & {\textbf{0.915}} &{0.733} & 0.810 & 0.799 & \textbf{0.868} & 0.816\\
    \texttt{text-davinci-003} & 0.862 & {0.814}  & {\textbf{0.938}} & \textbf{0.933} & 0.655 & \textbf{0.672} & 0.644 & 0.655\\
    \texttt{text-davinci-002} & 0.803 & {0.587} &{\textbf{0.861}}& 0.822  & 0.839 & \textbf{0.874} & 0.810 & 0.770\\
  \bottomrule
\end{tabular}
  \caption{Accuracy for prompt (P\textit{n}) and model combinations on the Civil Comments and GitHub Discussion datasets. The best-performing prompt(s) for each LLM API are highlighted in bold. We observed similar results for F1 scores. %
 }
  \label{tab:acc}
    \vspace{-15pt}
\end{table*}

\subsection{Observations}

\subsubsection{Prompt performance can regress over API updates.}
First of all, we found that regression does exist over API updates: 58.8\% of prompt + model combinations drop accuracy over API updates (Table~\ref{tab:acc}).
Among them, 70.2\% drop accuracy greater than 5\%.
Noticeably, \hl{across all different prompts,} the model update from \texttt{text-davinci-002} to \texttt{text-davinci-003} causes a consistent performance drop (16.8\% on average) on the GitHub Discussion Dataset but a consistent performance increase (11.8\% on average) on the Civil Comments dataset.
We hypothesize that the huge performance differences are due to the new training method \texttt{text-davinci-003} used, which causes major inconsistency across the two versions.

\subsubsection{Model updates affect different prompting strategies differently.}
We observed that among all model updates, 55\% do not cause a consistent performance drop or increase across prompts, i.e., the same model update helps some prompts but hurts others for the same task.
Specifically, we found that the simplest prompt, P1, drops accuracy in 75\% of the model updates, while the few-shot prompt, P4, only drops accuracy 45\% of all times.
Zooming in, we can see that the update from \texttt{gpt-3.5-turbo-0301} to \texttt{gpt-3.5-turbo-0613} caused a 9.6\% accuracy drop for P1, but a 5.1\% increase for P4 on the Civil Comments dataset.
This is particularly concerning, as the update is silent when a developer uses the main API \texttt{gpt-3.5-turbo}, which updates the underlying model from time to time.

Such non-uniform performance changes cause a major problem for prompt engineering: The developer may find that their carefully engineered prompt is no longer the best choice after a silent API update.
For example, the detailed instruction prompt (P3) has been the best-performing prompt up to the last model update,
but falls behind the few-shot prompt (P4) by 8.7\% on the latest model (\texttt{gpt-3.5-turbo-instruct}).
This indicates that prompt engineering is not a one-time effort, and calls for prompt versioning and prompt monitoring (see detailed discussion in Section~\ref{sec:prompt-ver}).

\subsubsection{Regressions happen even when prompt performance improves.}
We also found that overall 10.9\% individual predictions regress (from correct to wrong) over API updates. Almost always (87.9\%) when overall accuracy improves in an update, at least one previously correct prediction regresses.
For example, the model update from \texttt{text-davinci-002} to \texttt{text-davinci-003} improves P3's accuracy on the Civil Comments dataset by 7.7\%, but 1.8\% of the previously correct predictions now fail.

As such regressions are invisible in the aggregated accuracy scores,
it would be particularly concerning if the improvements and regressions are not uniform--the prompt may work better on some data slices but worse on others, causing fairness implications even when overall accuracy stays stable or improves.

\begin{table*}
  \small
  \begin{tabular}{lrrrrrr}
    \toprule
    \multirow{2}{*}{\textbf{Model}}&\multicolumn{3}{c}{\textbf{Civil Comments}} & \multicolumn{3}{c}{\textbf{GitHub Discussion}} \\    
    \cmidrule(lr){2-4}\cmidrule(lr){5-7}   
            &\textbf{Regression}&\textbf{Improvement}&\textbf{Unflipped} &\textbf{Regression}&\textbf{Improvement}&\textbf{Unflipped}

\\
    \midrule
    \texttt{gpt-3.5-turbo-instruct} & 0.319&0.289 &0.078 &0.186 &0.213 &0.258 \\
    \texttt{gpt-3.5-turbo-0613}  & 0.190& 0.138 & {0.025} & 0.267& 0.139 & 0.063 \\
    \texttt{gpt-3.5-turbo-0301} &0.075 & 0.096 & {0.006} &0.015 & {0.010} & 0.026 \\
    \texttt{text-davinci-003} &0.022 &0.028 &0.010 &0.005 &0.018 &0.018 \\
    \texttt{text-davinci-002} & 0.251 & 0.296 & {0.137} & 0.467 & 0.302 & {0.227}\\
    \texttt{average} & 0.171 & 0.169 & 0.051 & 0.188 & 0.136 & 0.118 \\
  \bottomrule
\end{tabular}
  \caption{Model entropy on the Civil Comments and GitHub Discussion datasets, averaged across all prompts.}
  \label{tab:entropy}
  \vspace{-10pt}
\end{table*}

\subsubsection{Regressions happen beyond the decision boundary.}
A natural hypothesis is that regressions happen on data points that models are less confident with (i.e. near the decision boundary). To explore this hypothesis, following existing work~\cite{zhang-etal-2019-mitigating}, we use information entropy to measure the model's confidence on a data point:
\[
    E_j = \sum_{i} -p_{ij}\cdot\log{p_{ij}}
\]
where $E_j$ is the entropy on input $j$, and $p_{ij}$ is the model's probability to predict label $i$ on input $j$.
Intuitively, when the model's prediction probabilities are more evenly distributed across different labels, the entropy is higher and the model is more uncertain on the input.
Since many LLM APIs do not expose the actual prediction probabilities, we approximate a model's prediction probabilities by running it on the same input multiple (n=20) times with a non-zero temperature (t=0.7).

Overall, we found that models are indeed more uncertain about flipping data points on average (Table~\ref{tab:entropy}). 
However, we also found that 63.8\% of regressions happen when models are very confident about their results (i.e. entropy $=0$).
This implies that model updates can drastically change predictions on data points far away from the decision boundary.

Across the models, we also found that different models show different levels of self-consistency: \texttt{gpt-3.5-turbo-0301} seems to be the most self-consistent one, while the update to \texttt{gpt-3.5-turbo-0613} makes it much less self-consistent.
This indicates another form of \emph{regression}: While the two models' accuracy is comparable, the update can affect model calibration~\cite{pmlr-v139-zhao21c} and make the model less self-consistent (or over-confident).

\begin{figure}
    \centering
    \includegraphics[width=0.9\linewidth]{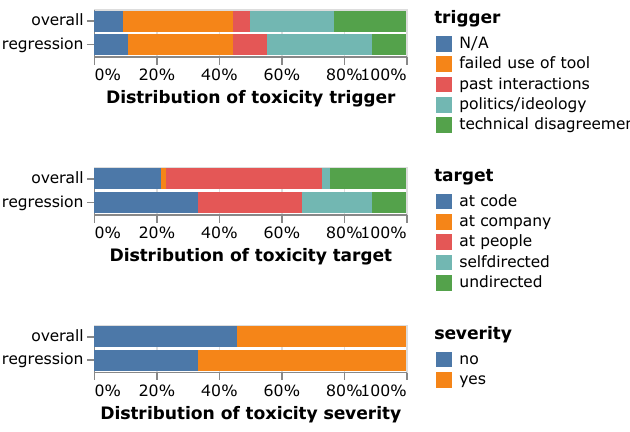}
    \vspace{-10pt}
    \caption{Regressions disproportionally happen when the toxicity relates to politics (25.7\% vs. 33.3\%), targets code (21.6\% vs. 33.3\%), or is severe (54.1\% vs. 66.7\%).}
    \label{fig:regression-slices}
    \vspace{-10pt}
\end{figure}

\subsubsection{Regressions are not uniform across data slices.}
We next explore where regressions happen systematically for specific data slices, with the metadata provided by the authors of the GitHub Discussion dataset~\cite{Github}.

We found that 90\% of regressions happen on toxic discussions, despite only 42.5\% of discussions being toxic in the dataset.
Breaking down the regression on toxic discussions by the provided metadata (Figure~\ref{fig:regression-slices}), 
we found that regressions are disproportionally common when the toxicity is triggered by politics (25.7\% \hl{overall} vs. 33.3\% \hl{among regressions}), targets code (21.6\% vs. 33.3\%), or is severe (54.1\% vs. 66.7\%),
suggesting that model updates can cause systematic worse performance for these specific data slices.

\subsubsection{Limitations.}
\hl{Readers should be careful when generalizing the results beyond the current experiment settings: We used specific prompt formats and sent the prompts as a single user request. The optimal prompt design may change when the LLM API varies.}

\section{Towards Regression Testing for Prompting LLMs}
Our exploratory case study highlights that model regression is a \textit{real} problem that is deeply affected by prompting and LLM non-determinism.
Based on our observations, we conclude with a discussion on how researchers can support regression testing for LLMs.

\subsection{Identifying Data Slices as Regression Test Suites}
In our case study, we found that individual predictions regress frequently (10.9\%).
Therefore, treating each data point as a regression test will simply be intractable.
An alternative would be to look at aggregated metrics over the entire dataset.
However, this level of monitoring is too coarse-grained and cannot inform developers on how to debug and adjust their prompts.

We argue that LLM regression tests should be at the level of \textit{slices}.
Our preliminary results show that it is possible to look at \textit{semantic} slices and localize where regressions happen (e.g., \textit{toxicity targeting code} for GitHub toxicity).
However, our slicing relies on extra metadata, which may not be available for many datasets.
Future research should further scaffold developers to identify data slices as regression test suites,
possibly by transferring existing approaches like slice discovery~\cite{eyuboglu2022domino} and error analysis~\cite{wu-etal-2019-errudite, zeno} on a single model to regression testing.

\subsection{Tracking Prompts for Regression Testing}
\label{sec:prompt-ver}
Our case study points out that prompt performance can be unstable across different APIs and each API has different best-performing prompts.
\hl{Therefore, developers need to track and update their prompt (possibly from a \textit{history} version), to maintain or improve prompt+LLM performance.}

However, existing prompt engineering practices provide insufficient support for prompt versioning and monitoring~\cite{johnnyprompt}--
Information and knowledge are often lost in the iterative prompt engineering process.
Future research can design systems for prompt+LLM tracking~\cite[e.g.,][]{modeltracker, mishra2023promptaid} to help developers explore behavioral changes, debug regressions, and update their prompts.

\subsection{Tackling Non-determinism in LLM Regression Testing}
Our case study shows that LLM predictions can flip a lot with a non-zero temperature.
This can cause lots of \textit{flakiness} when we perform regression testing for LLMs.
Future research on LLM regression testing should explicitly consider such non-determinism in their research design.
For example, to avoid a large sample size for each regression test, researchers can develop suitable statistical tests and test minimization strategies.

While our work focused on classification tasks, regressions can also happen for generative tasks,
where non-determinism is even more common for generating high-quality outputs.
To support regression testing LLMs on generative tasks, future research should consider incorporating multi-dimensional metrics~\cite{zhong-etal-2022-towards} and supporting developers in testing output properties specific to their requirements~\cite{ribeiro_2023}.

\begin{acks}
We thank Sherry Tongshuang Wu and Rohan Padhye for their discussion and feedback on this work.
\end{acks}

\bibliographystyle{ACM-Reference-Format}
\bibliography{sample-base}

\end{document}